\documentclass[dvips]{astavr}
\usepackage{supertabular,lscape,epsfig}
\usepackage{amssymb}
\usepackage{amsmath}

\SetPages{1}{6}
\SetVol{65}{2015}

\begin{document}

\begin{Titlepage}
\Title{Staying star gamma Equ }

\Author{V.D.~~B~y~c~h~k~o~v$^1$, L.V.~~B~y~c~h~k~o~v~a$^1$, J.~~M~a~d~e~j$^2$
          and~~G.~P.~T~o~p~i~l~s~k~a~y~a$^3$ }
{$^1$ Special Astrophysical Observatory of the Russian
        Academy of Sciences (SAO), Nizhnij Arkhyz, 369167 Russia \\
e-mail: (vbych,lbych)@sao.ru       \\
$^2$ Warsaw University Observatory, Al. Ujazdowskie 4, 00-478 Warszawa, Poland \\
e-mail: jm@astrouw.edu.pl          \\
$^3$  North Caucasian Federal University (NCFU), Pushkina 1, Stavropol, Russia \\
e-mail: gtop@mail.ru
}

\Received{Month Day, Year}
\end{Titlepage}

\Abstract{
We determined rotational period of practically nonrotating Ap star
gamma Equ and claim $P_{\rm rot}= 97$ years. This period is about 
35000 times larger than the average rotational period among stars 
of the same spectral class. Paper discusses possible mechanism 
explaining the origin of this phenomenon. }

{Stars: chemically peculiar -- Stars: magnetic fields -- Stars: individual: gamma Equ}

\section{Introduction}

All stars rotate from the birthtime, that is, from the stage of protostars,
to the last stages of evolution. Rotational speed plays an important role
in star formation stage and in further processes (Tassoul 1982). Accurate
value of the rotational period of a star can be determined if the star has
a spotty appearance of the surface, on which exist areas of temperature and
chemical inhomogeneities, as well as measureable magnetic fields. Such a
situation was found e.g. in Ap stars or in spotty BY Dra stars.

Gamma Equ is a well known magnetic Ap star. Magnetic field of that star
was studied from 68 years. Over the past 19 years longitudinal (effective)
magnetic field $B_e$ of $\gamma$ Equ was continously studied at the Special
Astrophysical Observatory (Russia). As the result, we collected the longest
and homogeneous series of measurements, consisting of total 130 $B_e$ points.
Such a unique record of homogeneous observations, which was obtained with
the same instruments and the same observers, allowed us to specify more
precisely the period of rotation and parameters of magnetic variability 
of that object. We determined, that both rotational and magnetic periods
of Gamma Equ, $P_{\rm rot} = P_{\rm mag} = 35394$ days. Note, that according
to the average statistical relationships period of rotation for stars of
this type should be equal to about 1 day (Bychkov et al. 2006; 2015).

\section{Analysis of observational data }

The average magnetic phase curve of $\gamma$ Equ was shown in Figure 1.
Based on that phase curve we derived the following parameters of the 
longitudinal magnetic field variations:

\vspace{1.5mm}\hbox{ \hspace{5mm} \vbox{
\hbox{$P = 35394.22 \pm 1171.5$ days $= 96.97 \pm 3.21$ years }
\hbox{$B_0 = - 266 \pm 6$ G }
\hbox{$B_1 = 851 \pm 8$ G   } } }

\noindent
which implies the following values of parameters, according to Stibbs-Preston
formalism (Stibbs 1950; Preston 1967):

\begin{itemize}
\item polar magnetic field strength, $B_p = 11400$ G,
\item surface magnetic field strength, $B_s = 7180$ G,
\item angle of inclination between the axis of rotation and the line
      of sight, $i = 63\deg $,
\item angle between the magnetic dipole axis and the axis of rotation,
      $\beta = 44\deg $.
\end{itemize}

We prewhitened the observed 68-yr $B_e$ time series from the long term
period and using standard methods of spectral analysis obtained shorter
period of the magnetic $B_e$ variability, $P = 6329.11 \pm 324.02$ days 
$= 17.34 \pm 0.89$ years with the amplitude of 120 G. The corresponing
phase curve with the period of 17.34 years is shown in Figure 2. This 
variability can be due to the precession of the rotation axis with that
period.

\begin{figure*}[h]
\includegraphics[angle=0, width=0.8\columnwidth]{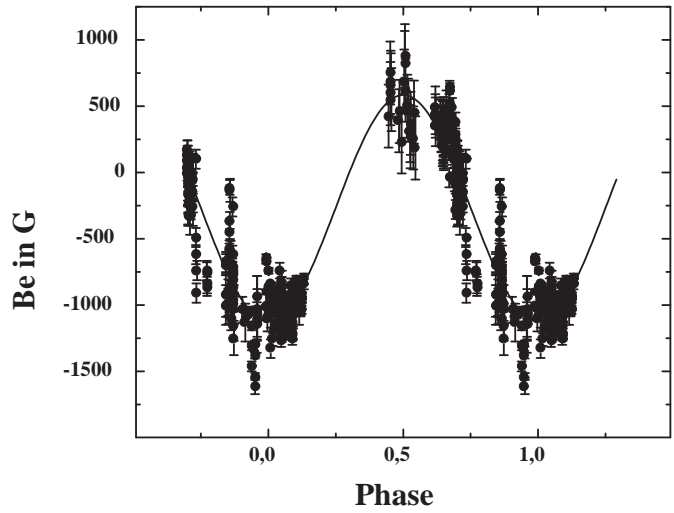}
\caption{Long-term magnetic phase curve for the rotational period
   of 97 years.}
\label{fig:1}
\end{figure*}

%\vspace{12mm}

\begin{figure*}[htb]
\includegraphics[angle=0, width=0.8\columnwidth]{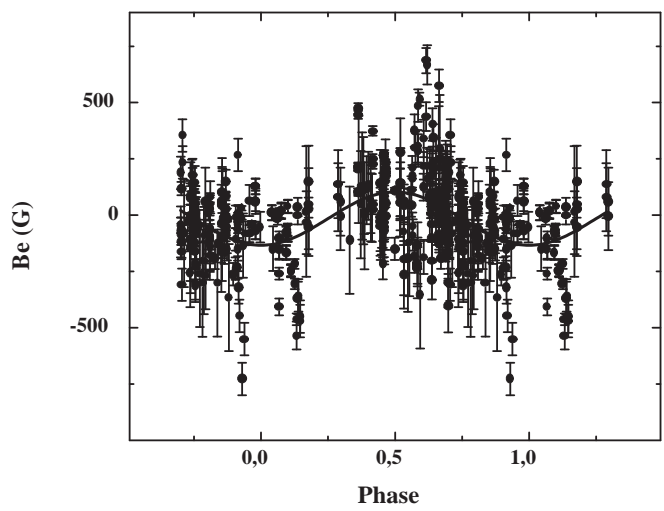}
\caption{Magnetic phase curve for the period of 17.3 years, which
   is not related to rotation. }
\label{fig:2}
\end{figure*}

\section{Discussion }

The most interesting problem is identification of the reason
why Ap star $\gamma$ Equ has incredibly long period of rotation. 
We believe that the very long period can be explained as due to
braking interaction between the global magnetic field of the
and surrounding circumstellar medium (Fabrika \& Bychkov 1988). 
Braking of the rotation is result of the "propeller" mechanism
(Illarionov \& Sunyaev 1975). As was pointed out by Fabrika \&
Bychkov (1988), propeller mechanism effectively works only if
$R_G > R_A > R_C$, wherein the capture radius $R_G$ is defined as:
\begin{equation}
R_{G} = \frac{2GM}{\xi^{2}v_{0}^{2}} \approx 2.0*10^{14}m_{z}\xi^{-2}v_{20}^{-2} cm
\end{equation}
alvenic radius, $R_A$ is
\begin{equation}
R_{A} = 2^{2/7}\mu^{4/7}\dot{M}_{a}^{-2/7}(2GM)^{-1/7} \approx 1.8*10^{14}\mu^{4/7}_{37}m_{s}^{-5/7}\xi^{6/7}v_{20}^{6/7} cm
\end{equation}
and corotation radius, $R_C$
\begin{equation}
R_{C} = {(\frac{GM}{\Omega^{2}})^{1/3}} \approx 6.71*10^{11}*P_{2}^{2/3}*m_{s}^{1/3} cm
\end{equation}

Change of the rotational period, $P$, proceeds according to the relation
\begin{equation}
P = \frac{P_{0}}{(1 - t/t_{a})}
\end{equation}
where $t_a$ equals (in years)
\begin{equation}
t_{a} =  4*10^{11}\mu^{-13/14}_{37}m_{s}^{-8/7}\hat{\eta}^{-6/7}_{1}\xi^{17/7}v_{20}^{18/7}
\end{equation}
Definition of variables were given in detail in Fabrika \& Bychkov (1988).

As can be seen from these equations, effectiveness of the "propeller"
mechanism is proportional to rotational period of the star and the 
density of surrounding medium. It turns out that relatively slowly 
rotating magnetic stars (slow rotators) can very effectively to lose 
angular momentum due to the interaction with the surrounding interstellar
medium under specified conditions. Using the estimates of fundamental 
physical parameters of $\gamma$ Equ published in Perraut et al. (2011)
and the formalism from Fabrika \& Bychkov (1988), we find that the star
$\gamma$ Equ could lose up to 90% of the total angular momentum during
time period of the order $10^7$ years. 

It should be noted that this process is not linear and depends on
a number of parameters. It should also be noted that slow magnetic 
rotators very effectively mix surrounding interstellar medium due 
to the action of "propeller" mechanism.

\section{Concluding remarks }

According to our opinion, such an unusually long period of rotation
of $\gamma$ Equ is the result of braking (loss of angular momentum) 
caused by the "propeller" mechanism (Illarionov \& Sunyaev 1975).
I.e. that braking results from interaction of the global magnetic 
field of $\gamma$ Equ with the interstellar medium. This star represents
probably the most striking example of a magnetic braking via
"propeller" mechanism. The process was currently recognized among 
magnetic Ap stars.

\Acknow{
Authors are grateful for the support of the Polish National Science
Center grant number 2011/03/B/ST9/03281, the Russian grant "Leading Scientific
School" and Presidential grants MK-6686.2013.2 and MK-1699.2014.2.
Research by Bychkov V.D. was supported by the
Russian Scientific Foundation grant N14-50-00043. }

\end{document}